\documentclass[aps,pre,twocolumn,groupedaddress]{revtex4}

\usepackage[dvips]{graphicx}
\begin{document}

\title{Solute trapping and diffusionless solidification in a binary system}

\author{Peter Galenko}
\email[]{e-mail: Peter.Galenko@dlr.de}
\affiliation{German Aerospace Center, Institute of Materials Physics in Space,
Cologne 51170, Germany}


\date{\today}
\begin{abstract}
Numerous experimental data on the rapid solidification of binary
systems exhibit the formation of metastable solid phases with the
initial (nominal) chemical composition. This fact is explained by
complete solute trapping leading to diffusionless (chemically
partitionless) solidification at a finite growth velocity of
crystals. Special attention is paid to developing a model of rapid
solidification which describes a transition from chemically
partitioned to diffusionless growth of crystals. Analytical
treatments lead to the condition for complete solute trapping which
directly follows from the analysis of the solute diffusion around
the solid-liquid interface and atomic attachment and detachment at
the interface. The resulting equations for the flux balance at the
interface take into account two kinetic parameters: diffusion speed
$V_{DI}$ on the interface and diffusion speed $V_D$ in bulk phases.
The model describes experimental data on nonequilibrium solute
partitioning in solidification of Si-As alloys [M.J. Aziz et al., J.
Cryst. Growth {\bf 148}, 172 (1995); Acta Mater. {\bf 48}, 4797
(2000)] for the whole range of solidification velocity investigated.
\\
\\
\begin{pacs}
\pacs{PACS numbers: 81.10.Aj; 05.70.Fh; 05.70.Ln; 81.30.Fb}
\end{pacs}

\end{abstract}



\keywords{Irreversible thermodynamics, phase transformation,
    diffuse interface, phase field}

\maketitle
\section{Introduction}\label{sec:intro}
The concept of ``solute trapping" has been introduced to define the
processes of solute redistribution at the interface which are
accompanied by $(i)$ the increasing of the chemical potential
\cite{bk1} and $(ii)$ the deviation of the partition coefficient for
solute distribution towards unity from its equilibrium value
(independently of the sign of the chemical potential) \cite{aziz4}.

In experimental investigations of rapid solidification, a complete
solute trapping leading to diffusionless (chemically partitionless)
solidification was first observed by Olsen and Hultgren and Duwez et
al. in experiments on rapid solidification \cite{oh91}. They showed
that rapidly solidifying alloy systems lead to the originating of
supersaturated solid solution with the initial (nominal) chemical
composition of the alloy. Later on, crystal microstructures with the
initial chemical composition were found by Biloni and Chalmers in
rapidly solidified pre-dendritic and dendritic patterns
\cite{bicha}.

Backer and Cahn \cite{bk1} have shown that with the finite
solidification velocity in a Cd-Zn system the coefficient of the Cd
distribution becomes equal to the unit that characterizes
diffusionless solidification. This fact has been confirmed in many
binary systems by Miroshnichenko \cite{mir}. He investigated
dendritic crystal microstructure after quenching from the liquid
state by splat quenching and melt spinning methods. The results of
Miroshnichenko's microstructural analysis show that at a cooling
rate greater than some critical value (depending on an alloy and
experimental method this value is in the range $10^5 - 10^6$ K/s) a
core of main stems of dendrites has initial (nominal) chemical
composition of the alloy. A critical value for undercooling in the
transition to purely thermally controlled growth with a homogeneous
distribution of chemical composition in Ni-B solidifying samples
processed by an electromagnetic levitation facility has been
obtained by Eckler et al. \cite{eck}. Finally, it is necessary to
note that many eutectic systems undergo chemically partitionless
solidification with an initial composition \cite{mir} that can be
explained by the transition to diffusionless solidification
\cite{gh}.

As a consequence, experimental investigations
\cite{bk1,oh91,bicha,mir,eck} show that with increasing driving
force of solidification solute traps are much more pronounced by
solidifying microstructure. At a finite value of the critical
governing parameters (undercooling, cooling rate or temperature
gradient) complete solute trapping occurs. Because the finite value
of the governing parameter defines the concrete solidification
velocity, complete solute trapping and diffusionless solidification
begin to proceed with a fixed critical growth velocity of crystals.

The main purpose of the present paper is to describe a model for
solute trapping and the transition from chemically partitioned to
diffusionless solidification in a binary system. Using the local
nonequilibrium approach to rapid solidification, an analysis of
diffusion mass transport in bulk phases together with conditions of
atomic attachment and detachment on the solid-liquid interface is
given.

The paper is organized as follows. In Sec. \ref{sec:prev}, previous
investigations of solute trapping are shortly reviewed. In Sec.
\ref{sec:dmass}, an analysis of solute diffusion leading to
pronounced solute trapping and complete solute trapping is given.
The nonequilibrium solute partitioning function for atoms on the
interface is derived in Sec. \ref{sec:spf}. A comparison with
previous models and experimental data on solidification of binary
systems is presented in Sec. \ref{sec:dexper}. Finally, in Sec.
\ref{sec:conls} conclusions of the work are summarized.

\section{Previous investigations}\label{sec:prev}
For the simplest case of an atomic system, let us consider an
isobaric and isothermal binary system (the pressure $P$ and
temperature $T$ are constant) with concentration $X_A$ and $X_B$ of
atoms $A$ and $B$, respectively. In this article, we denote $X$ as
the concentration of the atoms of $B$ sort. For a brief overview, we
summarize the equilibrium and nonequilibrium solute distribution on
the solid-liquid interface.
\subsection{Equilibrium}
In equilibrium, the concentration of atoms $X$ at the phase
interface is not equal from both sides of the interface due to the
different solubility of atoms in phases. During the equilibrium
coexistence of phases (gas-solid, liquid-solid, gas-liquid) the
atoms are distributed along the interface in consistency with the
diagram of a phase state. A difference in atomic concentration in
phases at the interface can be characterized by the equilibrium
coefficient $k_e$ of the atomic distribution between phases. For
equilibrium coexistence of phases (e.g., between crystal and melt,
vapor and crystal, crystal and liquid), the coefficient $k_e$ can be
expressed in the general form \cite{chernov1}
\begin{eqnarray}
k_e(X_L,X_S,T)=\frac{X_S^e}{X_L^e} \equiv \exp \left( - \frac{\Delta
\mu'}{RT} \right). \label{18}
\end{eqnarray}
In Eq.~(\ref{18}), $X_L^e$ and $X_S^e$ are the mole fractions of the
$B$ component in the liquid phase ($L$) or crystal ($S$),
respectively, $R$ is the gas constant, and $\Delta \mu'$ is the
difference in chemical potentials described by
\begin{eqnarray}
\Delta \mu'=\Delta \mu_B'-\Delta \mu_A',
\label{19}
\end{eqnarray}
with
\begin{eqnarray}
\Delta \mu_B'=\mu_{BS}'-\mu_{BL}', \qquad \Delta
\mu_A'=\mu_{AS}'-\mu_{AL}', \label{20a}
\end{eqnarray}
where $\Delta \mu_A'$ and $\Delta \mu_B'$ are the driving forces for
redistribution of atoms $A$ and $B$, respectively, which are defined
by redistribution potentials $\mu_A'$ and $\mu_B'$ for phases $L$
and $S$. The differences $\Delta \mu_A'$ and $\Delta \mu_B'$,
Eq.~(\ref{20a}), define the sign of $\Delta \mu'$ in Eq.~(\ref{19}).
For instance, if $\Delta \mu'$ is negative ($\Delta \mu_A'>\Delta
\mu_B'$), one has $k_e<1$ - the case of smaller solubility of atoms
$B$ in the phase $S$ in comparison with their solubility in the
phase $L$.

As a general characteristic of phase equilibria in binary systems,
expression (\ref{18}), together with Eqs.~(\ref{19}) and
(\ref{20a}), is usually considered as a measure of the driving force
for atomic redistribution at the phase interface. It can also be
considered as one of the main parameters for the construction of the
diagrams of a phase state.

\subsection{Nonequilibrium}
Expressions (\ref{18})-(\ref{20a}) assume local equilibrium at the
interface, which is a useful approximation for many systems
transforming at small interface velocities. At a large driving force
for the interface advancing and with increasing of the interface
velocity, the local equilibrium is not maintained \cite{bk1}.
Therefore, the condition for local interfacial equilibrium was
relaxed by taking into account a kinetic interface undercooling and
deviations from chemical equilibrium at the alloy's solidification
front \cite{chernov1,chernov}.

A number of models \cite{hall,chernov,aziz4,chernov2,brice,aziz3}
have been proposed to account for solute trapping and related
phenomena observed during rapid phase transformations. One of the
well-established boundary conditions for solute redistribution can
be taken from the continuous growth model (CGM) applied to solute
trapping by Aziz and Kaplan \cite{aziz4,aziz3,aziz5}. The CGM
assumes alloy solidification at a ``rough interface''; i.e, all
interface sites are potential sites for crystallization events. With
a high solidification rate, the atom can be trapped on a high-energy
site of the crystal lattice. This leads to a local nonequilibrium on
the interface and to the formation of metastable solids (see
examples in Ref. \cite{hghm}). As a result, the solute partitioning
function at the solid-liquid interface is described by
\cite{aziz3,aziz4}
\begin{equation}
k(V)= \frac{k_e+V/V_{DI}}{1+V/V_{DI}},
\label{28}
\end{equation}
where $V_{DI}$ is the speed of diffusion at the interface and $k_e$
is the value of the equilibrium partition coefficient given by
Eq.~(\ref{18}), i.e., with the negligible interface motion, $V
\rightarrow 0$. Equation (\ref{28}) evaluates the ratio $X_S/X_L$ at
the interface for dilute solutions of B (``solute'') in A
(``solvent'').

The interfacial diffusion speed $V_{DI}$ is the kinetic parameter
describing the deviation from chemical equilibrium at the interface.
It has been defined as the ratio between the diffusion coefficient
$D_I$ at the interface and the characteristic distance $\lambda$ for
the diffusion jump \cite{aziz4,aziz3}: $V_{DI}=D_I/\lambda$. The
distance $\lambda$ is assumed to be equal to the width of the
solid-liquid interface (few interatomic distances) and the diffusion
jumps are taken along the direction of growth. Therefore, this
definition for $V_{DI}$ is corrected by results of molecular dynamic
simulations \cite{cook}. They include diffusion in all spatial
directions; i.e., the diffusion speed is $V_{DI}=6D_I/\lambda$,
where the factor of 6 accounts for the possibility of jumps along
the six ($\pm x,y,z$) Cartesian axes.

Outcomes following from the solute partitioning function~(\ref{28})
were compared in the modeling of solute trapping using numerical
computations based on the phase-field theory of alloys
solidification. Wheeler et al. \cite{wbm2} naturally included an
energy penalty for high composition gradients in the liquid that
supresses the partitioning of solute at a rapidly moving interface
and leads to solute trapping. They also showed that the construction
of common tangents to the curves of free energy (in the spirit of
Baker and Cahn \cite{bk2}) has to be defined for nonequilibrium
concentrations which already depend on the solidification velocity.
In order to eliminate or reduce the solute trapping effect by the
diffuse interface at small growth velocity, Karma and co-workers
proposed an {\it ad hoc} suitable antitrapping condition to the
diffusion flux \cite{karma_sol}. These works \cite{wbm2,karma_sol}
showed that when the solute trapping effect comes to modeling alloy
solidification with both phase and concentration fields, a crucial
issue arises concerning the relative magnitudes of the gradients of
the two fields within the solidification front as well as the
relative thickness of the concentration jump interface.
Additionally, Conti \cite{conti} investigated the  usual
one-dimensional (1D) formulation of the phase field model without
the concentration gradient corrections of Wheeler et al.
\cite{wbm2}. He resolved the governing equations numerically for the
interface temperature and the solute concentration field as a
function of the growth velocity. The partition coefficient $k(V)$ is
monotonically increasing towards unity at large growth rates
following the predictions of the continuous growth model (\ref{28}).
However, in contrast to the results of natural experiments
\cite{bk1,oh91,bicha,mir,eck}, numeric predictions \cite{wbm2,conti}
were not able to reach the complete chemically partitionless
(diffusionless) solidification at a finite solidification velocity.

One of the deficiencies of the function (\ref{28}) is the difficulty
to describe complete solute trapping at the finite solidification
velocity: Equation (\ref{28}) predicts $k\rightarrow 1$ only with
$V\rightarrow \infty$. Contrary to this prediction, a transition to
partitionless solidification occurs at a finite solidification
velocity as it has been shown in numerous experiments
\cite{oh91,bicha,bk1,mir,eck}. Molecular dynamic simulations also
show that the transition to complete solute trapping is observed at
a finite interface velocity in rapid solidification of a binary
system \cite{cook}. Therefore, as an extension of Eq.~(\ref{28}), a
generalized function for solute partitioning in the case of local
nonequilibrium solute diffusion within the approximation of a dilute
system has been introduced by Sobolev \cite{S1}. This yields
\begin{eqnarray}
  & & k(V) = \frac{(1-V^2/V_D^2)k_e+V/V_{DI}}{1-V^2/V_D^2+V/V_{DI}},
  \qquad V<V_D, \nonumber \\
\nonumber \\
  & & k(V) = 1, \qquad V \geq V_D.
  \label{25}
\end{eqnarray}
The diffusion speed $V_{D}$ introduced in Eq. (\ref{25}) is the
characteristic bulk speed. It is defined as a maximum speed for the
solute diffusion propagation or as a speed for the front of solute
diffusion profile. In particular, the speed $V_D$ is obtained by the
speed of propagation of the plane harmonic wave away from the
solid-liquid interface (see the Appendix in Ref. \cite{G2}). As the
velocity $V$ of the interface is comparable by magnitude to the
speed $V_D$, the high frequency limit takes place:
$\omega\tau_D>>1$, where $\omega$ is the real cyclic frequency of
the plane harmonic wave and $\tau_D$ is the time for relaxation of
the diffusion flux to its steady state. In this case, $V_D$ has to
be considered finite and it is defined as $V_D=(D/\tau_D)^{1/2}$,
where $D$ is the diffusion coefficient in bulk liquid.

In the local equilibrium limit, i.e., when the bulk diffusive speed
is infinite, $V_D \rightarrow \infty$, expression (\ref{25}) reduces
to the function $k(V)$ which takes into account the deviation from
local equilibrium at the interface only as described by
Eq.~(\ref{28}). The function (\ref{25}) includes the deviation from
local equilibrium at the interface (introducing interfacial
diffusion speed $V_{DI}$) and in the bulk liquid (introducing
diffusive speed $V_D$ in the bulk liquid). As Eq. (\ref{25}) shows,
complete solute trapping $k(V)=1$ proceeds at $V=V_D$. This result
has been introduced by Sobolev from a postulation about the zero
value for the diffusion coefficient at $V\geq V_D$. The next section
further details that the condition for complete solute trapping
follows directly from the analysis of solute diffusion flux.

\section{Diffusion mass transport and solute trapping}\label{sec:dmass}
In 1D solidification along the $z$-axis, the mass balance is given by
\begin{eqnarray}
\frac{\partial X}{\partial t} = -\frac{\partial J}{\partial z},
\label{B511}
\end{eqnarray}
where $t$ is the time and $J$ is the diffusion flux. To consider
solute trapping in 1D local nonequilibrium solidification, we take
one of the results from a model of rapid phase transitions
\cite{gj}. Using this model, the evolution equation for diffusion
flux $J$ along the z-axis is described by
\begin{equation}
J=M\left [ \frac{\partial}{\partial z} \left ( \frac {\partial
s}{\partial X} + \varepsilon_x^2 \frac{\partial^2 X}{\partial z^2}
\right) - \alpha_j \frac {\partial J}{\partial t} \right],
\label{sol01}
\end{equation}
where $s$ is the entropy density, $\varepsilon_x$ the factor
proportional to the correlation length, and $M$ is the diffusion
mobility of atoms. The latter is defined by
\begin{equation}
M=T\overline D, \qquad \overline D=D(\partial (\Delta \mu)/\partial
X)^{-1}, \label{th00}
\end{equation}
where $D$ is the diffusion coefficient and $\Delta \mu$ is the
difference of chemical potentials between solvent and solute. From
known thermodynamic expressions \cite{j} one can accept that
\begin{equation}
\frac{\partial s}{\partial X}=-\frac{\Delta \mu}{T}, \qquad \alpha_j
= \frac{\tau_D}{TD}\frac{\partial (\Delta \mu )}{\partial
X}=\frac{\tau_D}{T\overline D}, \label{th01}
\end{equation}
where $\tau_D$ is the time for diffusion flux relaxation to its
steady state. Then, omitting the term responsible for atomic
correlation, i.e., assuming that $\varepsilon_x=0$, one can get from
Eq. (\ref{sol01}) the expression
\begin{equation}
J=M \left [ \frac{1}{T}\frac{\partial (\Delta \mu )}{\partial z} -
\frac{\tau_D}{T\overline D} \frac {\partial J}{\partial t} \right].
\label{sol02}
\end{equation}
Using Eq. (\ref{th00}), the evolution equation (\ref{sol02}) results
as follows
\begin{eqnarray}
\qquad J=-\overline D \frac{\partial (\Delta \mu )}{\partial z} -
\tau_D \frac{\partial J}{\partial t}. \label{sol03}
\end{eqnarray}

We find the solution for the diffusion flux $J$ which has
significance in the analysis of solute trapping. Using the
expression for the diffusion speed, $V_D=(D/\tau_D)^{1/2}$, from
Eqs. (\ref{B511}) and (\ref{sol03}) one gets
\begin{eqnarray}
\tau_D \frac{\partial^2 J}{\partial t^2}+\frac{\partial J}{\partial
t}= D \frac{\partial^2 J}{\partial z^2}. \label{sol100}
\end{eqnarray}
Equation (\ref{sol100}) is a partial differential equation of
hyperbolic type. It describes the flux $J$ in the so-called
``hyperbolic evolution'', which proceeds with a sharp front of the
profile for the solute transport. It occurs due to both the
diffusive and propagative nature of the transport in the high
frequency limit $\omega\tau_D>>1$ with $V\sim V_D$.

For a steady-state regime of interfacial motion  Eq. (\ref{sol100})
takes the form
\begin{eqnarray}
D\left(1-\frac{V^2}{V_D^2} \right) \frac{d^2J}{dz^2}+V\frac{dJ}{dz}=0,
\label{sol101}
\end{eqnarray}
which is true in a reference frame moving at constant velocity $V$
with the interface $z=0$. A general solution of Eq. (\ref{sol101})
is
\begin{eqnarray}
J(z)=c_1+c_2\exp \left(-\frac{Vz}{D(1-V^2/V_D^2)} \right).
\label{sol102}
\end{eqnarray}
To define a particular solution one can assume the following
boundary conditions: the balance on the interface is
$J(z=0)=V(X_L^*-X_S^*)$, and the flux is limited by the expression
$J(z\rightarrow \infty)=0$ far from the interface with $z\rightarrow
\infty$. The latter condition gives $c_1=0$ for any velocity $V$ and
one gets $c_2=0$ for $V\geq V_D$. Also, from the interfacial balance
with $z=0$ one gets $c_2=V(X_L^*-X_S^*)$ for $V<V_D$. As a result,
solution (\ref{sol102}) transforms into the particular solution
\begin{eqnarray}
& & J(z) \nonumber\\
& & =V(X_L^*-X_S^*)\exp \left(-\frac{Vz}{D(1-V^2/V_D^2)} \right), \qquad V<V_D, \nonumber\\
& & J(z)=0, \qquad V\geq V_D, \label{sol103}
\end{eqnarray}
where $X_L^*$ and $X_S^*$ are the liquid concentration and solid
concentration, respectively, on the interface.

Solution (\ref{sol103}) gives the condition for complete solute
trapping with finite velocity $V\geq V_D$. This is expressed by the
expression for the solute partitioning function
\begin{eqnarray}
& & k(V)=X_S^*/X_L^*\neq 1, \qquad V<V_D, \nonumber\\
& & k(V)=1, \qquad X_S^*=X_L^*, \qquad V\geq V_D. \label{sol104}
\end{eqnarray}
The latter condition in Eq. (\ref{sol104}) defines the equality of
the concentrations in the phases and leads to complete solute
trapping.

To obtain an explicit form for the solute partitioning function
(\ref{sol104}), we analyze the balance of diffusion fluxes on the
interface. Taking again the steady state regime of solidification
constant velocity $V$, from the system (\ref{B511})  and
(\ref{sol03}) one can obtain the equations
\begin{eqnarray}
\frac{dJ}{dz}=V\frac{dX}{dz}, \qquad J=-\overline D \frac{d(\Delta
\mu )}{dz} + \tau_D V\frac{dJ}{dz}, \label{sol04}
\end{eqnarray}
from which we get the single equation for the diffusion flux $J$.
This yields
\begin{eqnarray}
J=- D \left(\frac{\partial (\Delta \mu )}{\partial X}\right)^{-1}
\frac{d(\Delta \mu )}{dz} + D \frac{V^2}{V_D^2}\frac{dX}{dz}.
\label{sol05}
\end{eqnarray}
The above-defined thermodynamic parameter $\overline D$ and the
diffusion speed $V_D$ in bulk have been taken into account. Defining
the gradient of the difference of the chemical potentials as
$d(\Delta \mu)/dz = [\partial (\Delta \mu)/\partial X]dX/dz$, Eq.
(\ref{sol05}) gives
\begin{eqnarray}
J=- D \left(\frac{\partial (\Delta \mu )}{\partial X}\right)^{-1}
\left[1- \frac{V^2}{V_D^2} \right] \frac{d(\Delta \mu )}{dz}.
\label{sol06}
\end{eqnarray}
This equation is a general expression for the steady diffusion flux
into the liquid from the interface. Within the local equilibrium
limit $V_D\rightarrow \infty$ one can obtain the known Fickian
approximation which has been used previously for analysis of solute
trapping \cite{Kurz-Fisher,ahmad}.

Analytical solutions \cite{GS} for solidification under local
nonequilibrium diffusion show that the concentration in both phases
becomes equal to the initial (nominal) concentration and the
diffusion flux is absent for $V\geq V_D$. It is also given by Eq.
(\ref{sol103}). Therefore, in addition to Eq. (\ref{sol06}), one can
finally obtain
\begin{eqnarray}
& & J=- D \left(\frac{\partial (\Delta \mu )}{\partial
X}\right)^{-1} \left[1- \frac{V^2}{V_D^2} \right] \frac{d(\Delta \mu
)}{dz}, \qquad
V<V_D, \nonumber\\
\nonumber\\
& & J=0, \qquad V\geq V_D. \label{sol07}
\end{eqnarray}

At the phase interface one assumes in Eq. (\ref{sol07}), first, that
the term $D((\partial \Delta \mu )/\partial X)^{-1}$ is proportional
to concentration such that
\begin{eqnarray} D\left(\frac{\partial (\Delta \mu )}{\partial X}\right)^{-1}=\frac{D_IX_L^*}{RT}.
\label{sol08}
\end{eqnarray}
Second, the chemical inhomogeneity (solutal segregation) exists due
to the jump of the chemical potential $\Delta \mu$ which has the
interfacial gradient $-d(\Delta \mu )/dz\cong \Delta \mu/W_0$ at a
small distance $W_0$ of the order of a few interatomic distances.
Third, in an approximation of ideal (or even real) solutions, one
can assume for the interfacial difference of chemical potentials
$\Delta \mu =\Delta \mu^L(X)-\Delta \mu^S(X) \cong RT(\ln X_L^e -
\ln X_L^*) - RT(\ln X_S^e - \ln X_S^*) = RT\ln (X_S^*/X_L^*)- RT\ln
(X_S^e/X_L^e)=RT[\ln k(V) - \ln k_e]$, where $X_L^e$ and $X_S^e$ are
equilibrium concentrations on the interface from the liquid phase
and solid phase, respectively, and $k(V)$ is the ratio of
concentrations on the interface defined by Eq. (\ref{sol104}).
Taking into account these last evaluations, one can get for the
chemical potential gradient the expression
\begin{eqnarray}
- \frac{d(\Delta \mu )}{dz}\cong \frac{RT}{W_0}\ln \frac{k(V)}{k_e}.
\label{sol09}
\end{eqnarray}

Integration of the balance (\ref{sol04}) on the interface gives the
flux
\begin{eqnarray}
J=V(X_L^*-X_S^*). \label{sol10}
\end{eqnarray}
Substituting Eqs. (\ref{sol08}) and (\ref{sol09}) into the
expression for diffusion flux Eq. (\ref{sol07}) with using the
balance (\ref{sol10}) gives the following expression for the solute
partitioning function:
\begin{eqnarray}
& & \left[1- \frac{V^2}{V_D^2} \right]\ln \frac{k(V)}{ k_e}
=\frac{V}{V_{DI}}[1-k(V)], \qquad V<V_D, \nonumber\\
\nonumber\\
& & k(V)\equiv X_S^*/X_L^*=1, \qquad V\geq V_D, \label{sol11}
\end{eqnarray}
where $V_{DI}=D_I/W_0$ is the speed for solute diffusion on the
interface.

Equation (\ref{sol11}) gives the evaluation of solute trapping
effect through the solute partitioning function $k(V)$ derived
initially from the analysis of the evolution equation (\ref{sol01})
for the diffusion flux $J$. This equation takes into account finite
diffusion speeds on the interface and in bulk liquid. The
introduction of these two speeds is a consequence of the local
nonequilibrium both on the interface and in bulk liquid. As Eq.
(\ref{sol11}) shows, the complete solute trapping  $k(V)=1$ proceeds
at $V=V_D$. Equation (\ref{sol11}) transforms into a previously
known expression for the function $k(V)$ derived in Refs.
\cite{Kurz-Fisher,ahmad} with relaxing local equilibrium on the
interface and using local equilibrium in bulk liquid ($V_D
\rightarrow \infty$) for the diluted binary system ($X_L^*<<1$).

\section{Solute partitioning function}\label{sec:spf}
We use a model of diffusion in which particles move by diffusion
jumps in random time between two phases (states). This model was
called the ``two-level model of diffusion'' and it was introduced in
the context of various application, e.g., in chromatography
\cite{ge} or for a longitudinal solute dispersion in a tube with
flowing water (Taylor's dispersion) \cite{tay1}.

Let $P_i(t,z)$ be the probability density of a particle position in
the phase $i=L$ or in the phase $i=S$ at the moment $t$. Then local
conservation of the probability density in a point with coordinate
$z$ belonging to the phase $i$ is defined by
\begin{eqnarray}
\frac{\partial P_i}{\partial t}=-\frac{\partial J_i}{\partial z}.
\label{sol12}
\end{eqnarray}
If the interface moves with a velocity comparable to the solute
diffusion speed $V_D$ in bulk phases, then the flux $J_i(t,z)$ of
the density probability depends on the prehistory of the diffusion
process. The flux, therefore, is defined by
\begin{eqnarray}
J_i(t,z)=-\int^{t}_{-\infty} D_i(t-t^*)\frac{\partial
P_i(t^*,z)}{\partial z}dt^*. \label{sol13}
\end{eqnarray}

The relaxation function $D_i(t-t^*)$ can be chosen in the form
$D_i(t-t^*)=D_i(0)\exp [-(t-t^*)/\tau_D]$ of exponential decay. In
such a case, Eq. (\ref{sol13}) is reduced to the Maxwell-Kattaneo
equation
\begin{eqnarray}
\tau_D \frac{\partial J_i}{\partial t}+J_i+D_i(0)\frac{\partial
P_i}{\partial z}=0. \label{sol15}
\end{eqnarray}
It is accepted in Eq. (\ref{sol15}) that $D_i(0)$ is the diffusion
coefficient at the final moment of relaxation prehistory so that
$D_i(0)=D$. System (\ref{sol12}) and (\ref{sol15}) gives a single
equation of a hyperbolic type for the density of probability
\begin{eqnarray}
\tau_D \frac{\partial^2 P_i}{\partial t^2}+\frac{\partial
P_i}{\partial t} = D_i\frac{\partial^2 P_i}{\partial z^2},
\label{sol16}
\end{eqnarray}
or for the flux,
\begin{eqnarray}
\tau_D \frac{\partial^2 J_i}{\partial t^2}+\frac{\partial
J_i}{\partial t} = D_i\frac{\partial^2 J_i}{\partial z^2}.
\label{sol17}
\end{eqnarray}
As was shown in Ref. \cite{cz}, the density of probability described
by Eq. (\ref{sol16}) gives a positive entropy production for the
particle exchange between two levels (between two subsystems or
phases).

Integration of Eq. (\ref{sol16}) by an infinitesimal layer including
an interface leads to the balance
\begin{eqnarray}
\left[ D_i\frac{\partial P_i}{\partial z} +\tau_D \frac{\partial
(VP_i)}{\partial t} +J_i\right]\bigg|^{L}_{S}=0. \label{sol18}
\end{eqnarray}
In the steady-state regime one can get the following equation for
the i-th phase
\begin{eqnarray}
& & D_i\frac{\partial P_i}{\partial z} +\tau_D \frac{\partial
(VP_i)}{\partial t}
+J_i=D_i\frac{dP_i}{dz}-\tau_DV^2\frac{dP_i}{dz}+J_i\nonumber\\
\nonumber\\
& & =D_i\left(1-\frac{V^2}{V_{Di}^2}\right)\frac{dP_i}{dz}+J_i,
\label{sol19}
\end{eqnarray}
which is true in a reference frame moving with constant velocity $V$
and placed on the interface where the balance (\ref{sol12}) is
described as $dJ_i/dz=VdP_i/dz$. Using Eq. (\ref{sol19}), the
balance (\ref{sol18}) is
\begin{eqnarray}
& & J_L-J_S \nonumber\\
\nonumber\\
& & =-\left[D_L\left(1-\frac{V^2}{V_{DL}^2}\right)\frac{dP_L}{dz}
-D_S\left(1-\frac{V^2}{V_{DS}^2}\right)\frac{dP_S}{dz} \right].
\nonumber\\
\label{sol20}
\end{eqnarray}

In Eq. (\ref{sol20}) we introduce the speeds $V_{DL}$ and $V_{DS}$
of interfacial solute diffusion from the liquid and solid phases,
respectively. They are defined by
\begin{eqnarray}
V_{DL}=D_L/l_D=\nu_Ll_D, \qquad V_{DS}=D_S/l_D=\nu_Sl_D,
\label{sol20v}
\end{eqnarray}
where $D_L$ and $D_S$ are the diffusion coefficients in the phases,
$l_D$ scales for diffusion within which the diffusion jumps occur in
phases (or on the interface), and $\nu_L$ and $\nu_S$ are the
frequencies of diffusion jumps in phases (or on the interface). From
the theory of the transitive state \cite{christ} one can define the
frequencies of atomic jumps as
\begin{eqnarray}
& & \nu_L = \nu_0\exp \left( - \frac{Q_D}{RT} \right), \nonumber\\
\nonumber\\
& & \nu_S=\nu_0\exp \left( - \frac{Q_D+\Delta \mu'}{RT} \right), \label{8sl}
\end{eqnarray}
where $\nu_0$ is the attempt frequency of atomic jumps of the order
of the vibrational frequency \cite{chernov1,vin}, $Q_D$ the
activation barrier for atomic diffusion through the interface, and
$\Delta \mu'$ is the difference of chemical potentials defined by
Eqs. (\ref{19}) and (\ref{20a}). Obviously, interfacial equilibrium
exists for
\begin{eqnarray}
\nu_S/\nu_L=\exp [-\Delta \mu'/(RT)]\equiv k_e. \label{88sl}
\end{eqnarray}

From the interfacial balance (\ref{sol20}) there follows
\begin{eqnarray}
J_L-J_S=-\left(1-\frac{V^2}{V_{DL}^2}\right)\frac{d}{dz}
\left[D_LP_L-D_S(V)P_S\right], \label{sol21}
\end{eqnarray}
where $D_S(V)$ is the function of the interfacial velocity $V$
defined by
\begin{eqnarray}
& & D_S(V)=D_S\frac{1-V^2/V_{DS}^2}{1-V^2/V_{DL}^2} \nonumber\\
\nonumber\\
& & = \left\{
\begin{array}{ll}
0, &  V_{DS}<< V_{DL}, \\\\
D_S, &  V_{DS}\approx V_{DL}, \\\\
D_S(1-V^2/V_{DL}^2)^{-1}, & V_{DS}\rightarrow \infty.
\end{array}
\right. \label{sol22}
\end{eqnarray}
The function (\ref{sol22}) describes the following cases: (a)
$V_{DS}\rightarrow 0$, negligible diffusion in solid ($D_S=0$) in
comparison with the diffusion in liquid; (b) $V_{DS}\approx V_{DL}$,
approximate equality for diffusion speeds in the liquid and solid
around the interface; and (c) $V_{DS}\rightarrow \infty$, condition
of local equilibrium in the diffusion field of the solid (that
occurs with high frequency jumps of atoms in solid).

From now on, the above case (b) for approximate equality of
diffusion speeds in phases around the interface is taken. First, we
use the finite difference $-dx=l_D$ in the balance (\ref{sol20}).
Second, we take into account that the factor $(1-V^2/V_{DL}^2)$ is
related to the bulk diffusion. Finally, using the definition
(\ref{sol20v}), the balance (\ref{sol20}) is described by
\begin{eqnarray}
J_L-J_S=\left(1-\frac{V^2}{V_{D}^2}\right)[V_{DL}P_L-V_{DS}P_S],
\label{sol20a}
\end{eqnarray}
where $V_D$ is the solute diffusion speed in bulk liquid around the
interface. Using Eqs. (\ref{sol20v})-(\ref{88sl}), this balance can
be rewritten as
\begin{eqnarray}
J_L-J_S=V_{DL}\left(1-\frac{V^2}{V_D^2}\right)[P_L-k_eP_S].
\label{sol20b}
\end{eqnarray}

For the concentrated binary system the probabilities $P_L$ and $P_S$
in Eq. (\ref{sol20b}) are directly proportional to the atomic
concentrations in phases. This leads to
\begin{eqnarray}
P_L = X_S(1-X_L)/\Omega, \quad P_S = X_L(1-X_S)/\Omega, \label{9sl}
\end{eqnarray}
where $\Omega$ is the atomic volume. Therefore, Eq. (\ref{sol20b})
can be rewritten as
\begin{eqnarray}
& & J_L-J_S=\left(1-\frac{V^2}{V_D^2}\right) \nonumber\\
\nonumber\\
& & \times [X_S(1-X_L)-k_eX_L(1-X_S)]\frac{V_{DL}}{\Omega}.
\label{sol20u}
\end{eqnarray}
We further use the already obtained result (\ref{sol07}) according
to which the diffusion flux is absent at $V\geq V_D$. Then, the
difference (\ref{sol20u}) of fluxes on the interface takes the form
\begin{eqnarray}
& & J_L-J_S = \left(1-\frac{V^2}{V_D^2}\right)
\nonumber\\
\nonumber\\
& & \times [X_S(1-X_L)-k_eX_L(1-X_S)]\frac{V_{DL}}{\Omega},
\qquad V<V_D, \nonumber\\
\nonumber\\
& & J_L=J_S, \qquad V\geq V_D. \label{sol20d}
\end{eqnarray}

The net flux (\ref{sol20d}) must be equal to the diffusion flux
\begin{eqnarray}
J_D = (X_L-X_S) \frac{V}{\Omega}. \label{21s}
\end{eqnarray}
From the equality of Eqs. (\ref{sol20d}) and (\ref{21s}) one gets
\begin{eqnarray}
& & (X_L - X_S) \frac{V}{V_{DI}}= \left(1-\frac{V^2}{V_D^2}
\right) \nonumber\\
\nonumber\\
& & \times [X_S(1-X_L)-k_eX_L(1-X_S)],
\qquad  V<V_D, \nonumber\\
 \nonumber\\
& & X_L=X_S, \qquad V \geq V_D, \label{22s}
\end{eqnarray}
in which $V_{DI}=V_{DL}$ is the diffusion speed on the interface
from the liquid phase. Equation (\ref{22s}) can be easily resolved
regarding the function $k(V)=X_S/X_L$ of nonequilibrium solute
partitioning. This yields
\begin{eqnarray}
& & k(V,X_L^*)  \nonumber\\
\nonumber\\
& & = \displaystyle
\frac{(1-V^2/V_D^2)k_e+V/V_{DI}}{(1-V^2/V_D^2)[1-(1-k_e)X_L^*]+V/V_{DI}},
\quad V<V_D, \nonumber\\
\nonumber\\
& & k(V,X_L^*=X_0)= 1, \qquad V \geq V_D, \label{23s}
\end{eqnarray}
where $X_L^*$ is the solute concentration in the liquid at the
interface.

\begin{figure}
\includegraphics[width=0.485\textwidth]{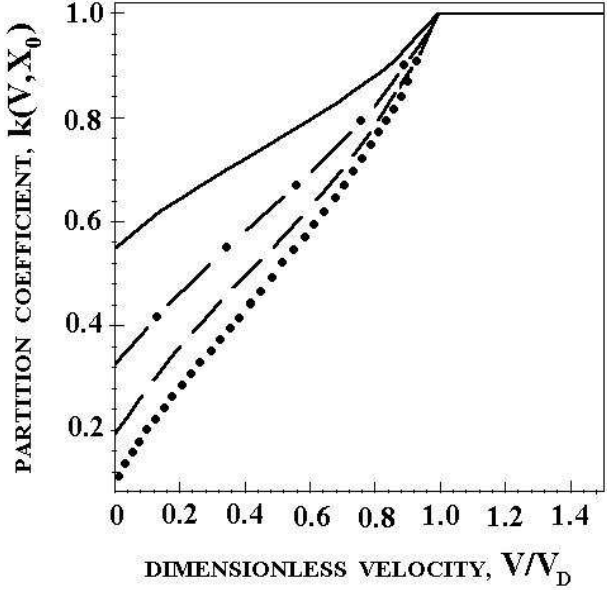}
\caption{Predictions of the model given by Eq.~(\ref{25A}).
Constants of the binary system are equilibrium partition coefficient
$k_e=0.1$, bulk diffusion speed $V_D=25$ (m/s), and interface
diffusion speed $V_{DI}=20$ (m/s). The curves present: diluted
system, $(1-k_e)X_0<<1$ (dotted line); slightly concentrated system,
$X_0=0.10$ (dashed line); concentrated system $X_0=0.25$,
(dash-dotted line); and equiconcentrated system, $X_0=0.50$ (solid
line). } \label{fig:g1}
\end{figure}

\begin{figure}
\includegraphics[width=0.485\textwidth]{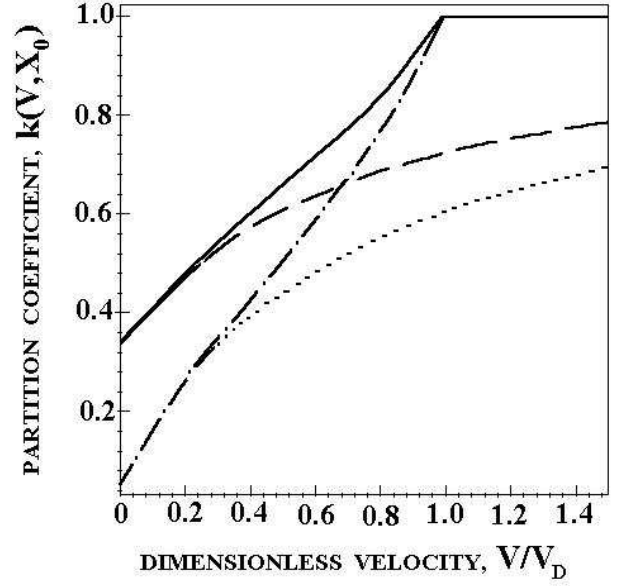}
\caption{Nonequilibrium solute partitioning function $k(V,X_0)$
given by the various models. Constants of the binary system are
nominal concentration of a solute $X_0=0.05$ mole fraction,
equilibrium partition coefficient $k_e=0.22$, bulk diffusion speed
$V_D=19$ (m/s), and interface diffusion speed $V_{DI}=16$ (m/s). The
dotted line is given by the model of Aziz \cite{aziz3} for the
diluted system $(1-k_e)X_0<<1$, the dashed line is given by the
model of Aziz and Kaplan \cite{aziz4}, the dash-dotted line is given
by the model of Sobolev \cite{S1} for diluted system, and the solid
line is predicted by the present model given by Eq.~(\ref{25A}).}
\label{fig:gasak} \end{figure}

\section{Discussion and comparison with experimental data}\label{sec:dexper}
Expression (\ref{23s}) gives the general functional dependence of
solute partitioning at the phase interface for concentrated binary
systems, and with application to rapid solidification, it exhibits
the two known limits. In the first limit, when solidification
proceeds with local nonequilibrium at the interface only, i.e., with
$V_D \rightarrow \infty$, Eq.~(\ref{23s}) leads to the solute
partitioning function of Aziz and Kaplan \cite{aziz4}. In the second
limit, as the concentration $X_L^*$ of the second dissolved
component becomes small, i.e., the term $(1-k_e)X_L^*$ might be
negligible in comparison with the unity, Eq.~(\ref{23s}) transforms
into Eq.~(\ref{25}) as suggested by Sobolev \cite{S1}.

From the analytical solution of the problem of rapid solidification
under the steady-state regime \cite{GS}, the concentration at the
planar interface is given by
 \begin{eqnarray}
& & X_L^*=\frac{X_0}{k(V)}, \qquad V<V_D, \nonumber\\
\nonumber\\
& & X_L^*=X_0, \qquad   V \geq V_D, \label{24}
\end{eqnarray}
where $X_0$ is the nominal (initial) concentration of the solute in
the system. In accordance with the solutions obtained in
Refs.~\cite{GS}, a source of concentration perturbations, i.e., the
solid-liquid interface, moving at a velocity $V$ equal to or higher
than the maximum speed $V_D$ of these perturbations, cannot change
the concentration or create the concentration profile ahead of
itself. As a result for the interface, one obtains in Eq.~(\ref{24})
that $X_L = X_S = X_0$ with $V \geq V_D$. Then the substitution of
Eq.~(\ref{24}) into Eq.~(\ref{23s}) leads to the following
expression for nonequilibrium solute partitioning function:
\begin{eqnarray}
& & k(V,X_0) \nonumber\\
& & = \frac{(1-V^2/V_D^2)[k_e + (1-k_e)X_0] + V/V_{DI}}{1-V^2/V_D^2+ V/V_{DI} }, V<V_D, \nonumber\\
\nonumber\\
& & k(V,X_0) = 1, \qquad    V \geq V_D.
\label{25A}
\end{eqnarray}

Figure \ref{fig:g1} demonstrates the behavior of solute
partitioning, Eq.~(\ref{25A}), as a function of the interface
velocity at various nominal solute concentrations. As the system
deviates from a diluted one, the trapping of a solute becomes much
more pronounced. Also, Eq.~(\ref{25A}) shows that, independently
from the solute concentration within the system, the complete solute
trapping $k(V,X_0)=1$ proceeds when the interface velocity becomes
equal to or greater than the diffusion speed, i.e., with $V \geq
V_D$. The condition of equality of concentrations in the liquid and
solid [see Eqs. (\ref{22s}) and (\ref{23s})] means that the lines of
the nonequilibrium kinetic liquidus and solidus in the kinetic phase
diagram are merging. It can also be considered as the
characteristics of diffusionless processes.

As a general outcome, Eq.~(\ref{25A}) includes the following
important cases for nonequilibrium phase transformations: ({\it i})
the dilute limit described by Aziz's model \cite{aziz3},
Eq.~(\ref{28}),
\begin{eqnarray}
(1-k_e)X_0<<1, ~~and~V_D \rightarrow \infty, \nonumber
\end{eqnarray}
({\it ii}) the dilute limit described by Sobolev's solute
partitioning function, Eq.~(\ref{25}),
\begin{eqnarray}
(1-k_e)X_0<<1,  ~~and~with~the~finite~V_D, \nonumber
\end{eqnarray}
({\it iii}) the concentrated system described by Aziz and Kaplan's
model, Ref.~\cite{aziz4},
\begin{eqnarray}
V_D \rightarrow \infty, ~~for~arbitrary~concentration~X_0.
\nonumber
\end{eqnarray}
In comparison with the present model's prediction described by
Eq.~(\ref{25A}) these limits are plotted in Fig.~\ref{fig:gasak}.

\begin{figure}
\includegraphics[width=0.495\textwidth]{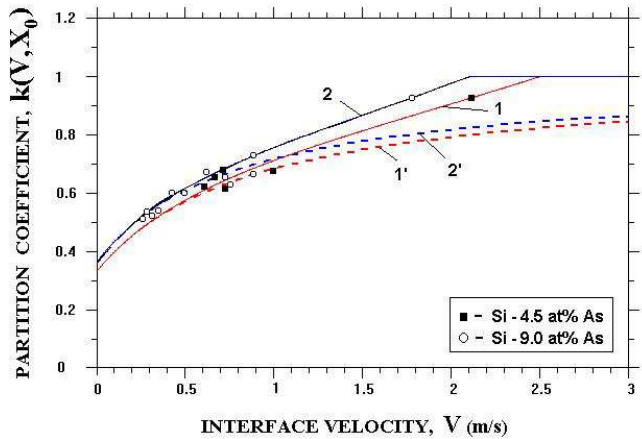}
\caption{Solute partitioning versus interface velocity for
experimental data \cite{aziz1,aziz2} on solidification of Si-As
alloys. Curves $1^\prime$ and $2^\prime$ are given by Eq.
(\ref{25A}) with $V_D\rightarrow \infty$ for 4.5 at.\% and 9.0 at.\%
of As in Si, respectively (that gives the model of Azis and Kaplan
\cite{aziz4}). They describe experiment at small and moderate
solidification velocities. Curves $1$ and $2$ are given by
(\ref{25A}) with the finite speed $V_D$ for 4.5 at.\% and 9.0 at.\%
of As in Si, respectively. These show ability to describe experiment
in a whole region of investigated solidification velocities for both
alloys. Data for calculations are given in Table \ref{tab:1}.}
\label{fig:Si_As}
\end{figure}

Figure \ref{fig:Si_As} exhibits theoretical predictions for solute
partitioning in comparison with experimental data on the
solidification of Si-As alloys. Introducing the deviation from
equilibrium at both the interface and bulk liquid allows one to
describe the whole set of experimental data. Particularly, the
complete solute trapping is predicted by Eq. (\ref{25A}) for Si-4.5
at.\%As with $V_D=2.5$ m/s and for Si-9.0 at.\%As with $V_D=2.1$ m/s
(Table \ref{tab:1}). This provides a much better agreement with
experiments than that shown by the Aziz-Kaplan model.

As can be seen in Fig. \ref{fig:Si_As}, predictions of the model of
Azis and Kaplan [Eq. (\ref{25A}) with $V_D\rightarrow \infty$]
disagree with experimental data in the region $1.7<V$ (m/s)$<2.2$ of
solidification velocities. One may note that at the same
solidification velocity, i.e., below about V = 2 (m/s), the
"interface temperature - velocity" relationship also exhibits a
clear deviation from experimental data (see Fig. 11 in Ref.
\cite{aziz2}). One may also attribute this deviation to the
increasing influence of local nonequilibrium solute diffusion around
the interface and intensive solute trapping. Thermodynamic analysis
and numeric evaluations confirm the idea about the pronounced
influence of local equilibrium in bulk liquid on solute trapping and
"interface temperature - velocity" relationship at high
solidification velocity \cite{G2,Gmater2}. This example confirms
that local nonequilibrium in the solute diffusion field is
responsible for nonequilibrium effects appearing in rapid
solidification (such as solute trapping and solute drag) and
essential influence on the interface response functions
(temperature, concentration, velocity) \cite{Gmater2}. Thus, the
agreement between Eq. (\ref{25A}) and experimental data demonstrates
the pronounced effect of deviation from local equilibrium in bulk
liquid on solute trapping at higher solidification velocity.

\begin{table*}
\caption{\label{tab:1} Interface diffusion speed $V_{DI}$ and bulk
diffusion speed $V_D$ for  binary systems used in the calculations
of the partitioning function $k(V,X_0)$ at the solid-liquid planar
interface. Equilibrium partition coefficient is taken as $k_e$=0.3
from Refs. \cite{aziz1,aziz2}.}
\begin{ruledtabular}
\begin{tabular}{lllll}
    Model    & Binary system     & $V_{DI}$~(m/s) & $V_D$~(m/s) & Reference \\
\hline
    Aziz and Kaplan's model, Ref.~\cite{aziz4} & Si - 4.5~at.\%~As & 0.46 & -- & \cite{aziz1}  \\
                         &      & 0.37 & -- & \cite{aziz2}  \\
                                \hline
    Aziz and Kaplan's model, Ref.~\cite{aziz4} & Si - 9~at.\%~As & 0.46 & -- & \cite{aziz1}  \\
                         &      & 0.37 & -- & \cite{aziz2}  \\
                                \hline
Sobolev's solute partitioning function, Eq.~(\ref{25}) & Si - 4.5~at.\%~As  & 0.75 & 2.7 & \cite{S1}  \\
                         &   and Si - 9~at.\%~As    &  &  &  \\
                \hline
    Present model, Eq.~(\ref{25A})  & Si - 4.5~at.\%~As & 0.8 & 2.5 & current data  \\
                    & Si - 9~at.\%~As & 0.8 & 2.1 & \cite{G2}  \\
\end{tabular}
\end{ruledtabular}
\end{table*}

Summarizing the behavior for solute partitioning shown in Figs.
\ref{fig:g1}-\ref{fig:Si_As}, one can conclude that during rapid
solidification the consequences of deviations from local chemical
equilibrium are threefold. First, the partition coefficient becomes
dependent on the growth velocity. Second, the liquidus and solidus
lines approach each other. For these two cases it can be enough to
introduce into the theory deviation from local equilibrium at the
interface only. Third, in the extreme case (if the solidification
velocity is equal to or greater than the atomic diffusive speed in
bulk liquid) the partition coefficient $k(V)$ becomes unity and the
liquidus and solidus lines coincide. This leads to a solid being far
from chemical equilibrium upon diffusionless solidification. Such
three conditions are of special importance in the preparation of
metastable supersaturated solutions \cite{hghm}.

\section{Conclusions}\label{sec:conls}
Solute trapping in rapid solidification of a binary alloy's system
has been considered. It has been shown that the condition for
complete solute trapping leading to diffusionless solidification
follows directly from the solution for the diffusion task. This task
assumes both the low-frequency regime (purely diffusion) and
high-frequency regime (diffusion and propagative regime) of atomic
motion in a phenomenological statement.

The two-level model has been used to define the solute partitioning
function. This model has been used previously (e.g., in
chromatography and for investigation of longitudinal solute
dispersion), and it has been formally reduced to expressions for an
extended version of the continuous growth model. The extended
version adopts two kinetic parameters: solute diffusion speed
$V_{DI}$ on the interface and solute diffusion speed $V_D$ in bulk
liquid.

A condition of complete solute trapping at the finite solidification
velocity equal to the diffusion speed, $V=V_D$, has been found. This
fact is expressed by the general expression (\ref{sol104}) for the
solute partitioning function. This condition defines the equality of
the concentration in the phases and describes complete solute
trapping. Analysis leads to concrete forms for the solute
partitioning function. The first function is given by Eq.
(\ref{sol11}) and the second function for solute partitioning is
described by Eq. (\ref{23s}). Both these functions predict a sharp
finishing of solute trapping and the onset of diffusionless crystal
growth at the solidification velocity $V$ equal to the solute
diffusion speed $V_D$ in bulk liquid. A concrete expression for the
liquid concentration $X_L^*$ at the interface allows us to give
predictions comparable with experimental data.

The model predicts the complete behavior for the solute partitioning
function dependent on the solidification velocity and alloy
concentration. In comparison with the experimental data of Aziz et
al. on solidification of Si-As alloys [M.J. Aziz et al., J. Cryst.
Growth {\bf 148}, 172 (1995); Acta Mater. {\bf 48}, 4797 (2000)],
the model well predicts deviation of the solute partitioning from
equilibrium and complete solute trapping (Fig. \ref{fig:Si_As}). The
transition from chemically partition growth to diffusionless growth
at $V=V_D$ occurs sharply. As has been shown for dendritic growth
\cite{GD1} such a sharp transition leads to an abrupt exchange of
growth kinetics in consistency with experimental data.

\begin{acknowledgments}
The author thanks Professor Dieter Herlach and Professor Dmitri
Temkin for numerous useful discussions. This work was performed with
support from the German Research Foundation (DFG - Deutsche
Forschungsgemeinschaft) under project No. HE 1601/13.
\end{acknowledgments}

\bibliography{basename of .bib file}

\begin{thebibliography}{99}
\bibitem{bk1}
J.C. Baker and J.W. Cahn, Acta Metall. {\bf 17}, 575 (1969); in {\em
Solidification\/}, edited by T.J. Hughel and G.F. Bolling (American
Society of Metals, Metals Park, OH, 1971) p.23.

\bibitem{aziz4}
M.J. Aziz and T. Kaplan, Acta Metall. {\bf 36}, 2335 (1988).

\bibitem{oh91}
W.T. Olsen and R. Hultgren, Trans. AIME, {\bf 188} 1323 (1950); P.
Duwez, R.H. Willens, and W. Klement (Jr.) J. Appl. Phys. {\bf 31},
1136 (1960).

\bibitem{bicha}
H. Biloni and B. Chalmers, Transactions AIME {\bf 233}, 373 (1965).

\bibitem{mir}
I.S. Miroshnichenko, {\em Quenching From the Liquid State\/}
(Metallurgia, Moscow, 1982).

\bibitem{eck}
K. Eckler, R.F. Cochrane, D.M. Herlach, B.  Feuerbacher, and M.
Jurisch, Phys. Rev. B {\bf 45}, 5019 (1992).

\bibitem{gh}
P.K. Galenko and D.M. Herlach, Phys. Rev. Lett. {\bf 96}, 150602
(2006).

\bibitem{chernov1}
A.A. Chernov, in {\em Modern  Crystallography\/}, edited by M.
Cardona, P. Fulde and H.-J. Queisser, {\em Springer Series in
Solid-State Science Vol.36\/} (Springer, Berlin, 1984), Vol.III,
Chap.4.

\bibitem{hall}
R.N. Hall, J. Phys. Chem. {\bf 57}, 836 (1953).

\bibitem{chernov}
A.A. Chernov, Sov. Phys. Uspekhi {\bf 13}, 101 (1970).

\bibitem{chernov2}
A.A. Chernov, in {\em Rost Kristallov\/}, edited by A.V. Shubnikov
and N.N. Sheftal, {\em Vol.3\/}, (Akademia Nauk SSSR, Moscow, 1959)
[{\em English translation: \/} "Growth of Crystals", vol.3
(Consultants Buro, New York, 1962) p.35]; V.V Voronkov and A.A.
Chernov, Sov. Phys. Crystallogr. {\bf 12}, 186 (1967).

\bibitem{brice}
J.C. Brice, {\em The Growth of Crystals from the Melt\/}
(North-Holland, Amsterdam, 1965), p. 65; K.A. Jackson, G.H. Gilmer,
and H.J. Leamy, in {\em Laser and Electron Processing of
Materials\/}, edited by C.W. White and P.C. Peercy (Academic Press,
New York, 1980), p. 104; R.F. Wood, Appl. Phys. Lett. {\bf 37}, 302
(1980); D.E. Temkin, Sov. Phys. Crystallogr. {\bf 32}(6), 782
(1988).

\bibitem{aziz3}
M.J. Aziz, J. Appl. Phys. {\bf 53}, 1158 (1982).

\bibitem{aziz5}
M.J. Aziz and W.J. Boettinger, Acta Metall. Mater. {\bf 42}, 527
(1994); M.J. Aziz, Metall. Mater. Trans {\bf 27A}, 671 (1996).

\bibitem{hghm}
D. Herlach, P. Galenko, and D. Holland-Moritz, {\em Metastable
Solids From Undercooled Melts\/} (Elsevier, Amsterdam, 2007).

\bibitem{cook}
S.J. Cook and P. Clancy, J. Chem. Phys. {\bf 99}, 2175 (1993).

\bibitem{wbm2}
A.A. Wheeler, W.J. Boettinger, and G.B. McFadden, Phys. Rev. E {\bf
47}, 1893 (1993); W.J. Boettinger, A.A. Wheeler, B.T. Murray, and
G.B. McFadden, Mater. Sci. Eng. A {\bf 178}, 217 (1994).

\bibitem{bk2}
J. C. Baker and J. W. Cahn, {\it Solidification} (ASM, Metals Park,
OH, 1971), p. 23.

\bibitem{karma_sol}
A. Karma, Phys. Rev. Lett. {\bf 87}, 115701 (2001); J.J. Hoyt, M.
Asta, and A. Karma, Mater. Sci. Eng. R {\bf 41}(6), 121 (2003); J.C.
Ramirez, C. Beckermann, A. Karma, and H.-J. Diepers, Phys. Rev. E
{\bf 69}, 051607 (2004); B. Echebarria, R. Folch, A. Karma, and M.
Plapp, Phys. Rev. E {\bf 70}, 061604 (2004).

\bibitem{conti}
M. Conti, Phys. Rev. E {\bf 56}, 3717 (1997).

\bibitem{S1}
S.L. Sobolev, Phys. Status Solidi A {\bf 156}, 293 (1996).

\bibitem{G2}
P. Galenko, Phys. Rev. B {\bf 65}, 144103 (2002).

\bibitem{gj}
P. Galenko and D. Jou, Phys. Rev. E {\bf 71}, 046125 (2005).

\bibitem{j}
D. Jou, J. Casas-Vazquez, and G. Lebon, {\it Extended Irreversible
Thermodynamics}, 2nd Edition (Springer, Berlin, 1996).

\bibitem{Kurz-Fisher}
W. Kurz and D.J. Fisher, {\it Fundamentals of Solidification}, 3rd
ed. (Trans Tech, Aedermannsdorf, 1992).

\bibitem{ahmad}
N.A. Ahmad, A.A. Wheeler, W.J. Boettinger, and G.B. McFadden, Phys.
Rev. E {\bf 58}, 3436 (1998).

\bibitem{GS}
P. Galenko and S. Sobolev, Phys. Rev. E {\bf 55}, 343 (1997); P.K.
Galenko and D.A. Danilov, J. Cryst. Growth {\bf 216}, 512 (2000);
Phys. Lett. A {\bf 272}, 207 (2000); Phys. Rev. E {\bf 69}, 051608
(2004).

\bibitem{ge}
J.C. Giddings and H. Eyring, J. Phys. Chem. {\bf 59}, 416 (1955);
J.C. Giddings, J. Chem. Phys. {\bf 26}, 169 (1955).

\bibitem{tay1}
G.I. Taylor, Proc. R. Soc. London, Ser. A {\bf 219}, 186 (1953);
ibidem {\bf 223}, 446 (1954); Van Den C. Broeck, Physica A {\bf
186}, 677 (1990).

\bibitem{cz}
J. Camacho and M. Zakari, Phys. Rev. E {\bf 50}, 4233 (1994).

\bibitem{christ}
J.W. Christian, {\it The Theory of Transformations in Metals and
Alloys}, 2nd Edition (Pergamon Press, Oxford, 1975), P. 1, Chapt. 3.

\bibitem{vin}
G.H. Vineyard, J. Phys. Chem. Solids {\bf 3}, 121 (1957).

\bibitem{aziz1}
J.A. Kittl, M.J. Aziz, D.P. Brunco, and M.O. Thompson,
J. Cryst. Growth {\bf 148}, 172 (1995).

\bibitem{aziz2}
J.A. Kittl, P.G. Sanders, M.J. Aziz, D.P. Brunco, and M.O. Thompson,
Acta Mater. {\bf 48}, 4797 (2000).

\bibitem{Gmater2}
P. Galenko, Mater. Sci. Eng. A {\bf 375-377}, 493 (2004).

\bibitem{GD1}
P.K. Galenko and D.A. Danilov, Phys. Lett. A {\bf 235}, 271 (1997);
J. Cryst. Growth {\bf 197}, 992 (1999).

\end{thebibliography}

\end{document}